\documentclass[letterpaper,twocolumn,superscriptaddress,floatfix]{revtex4}
\usepackage[latin1]{inputenc}
\usepackage{bm}
\usepackage{multirow,amssymb,amsbsy,amsmath}
\usepackage{graphicx}
\usepackage{verbatim}
\usepackage{pifont}
\usepackage{subfigure}
\usepackage{float}


\begin{document}

\title{Probing the collective dynamics of nuclear spin bath in a rare-earth
ion doped crystal }
\author{Zong-Quan Zhou}
\author{Yu Ma}
\author{Tao Tu$\footnote{email:tutao@ustc.edu.cn}$}
\author{Pei-Yun Li}
\author{Zong-Feng Li}
\author{Chao Liu}
\author{Peng-Jun Liang}
\author{Xiao Liu}
\author{Yi-Lin Hua}
\author{Tian-Shu Yang}
\author{Jun Hu}
\author{Xue Li}
\author{Yi-Xin Xiao}
\author{Yong-Jian Han}
\author{Chuan-Feng Li$\footnote{email:cfli@ustc.edu.cn}$}
\author{Guang-Can Guo}
\affiliation{CAS Key Laboratory of Quantum Information, University of Science and
Technology of China, CAS, Hefei, 230026, China}
\affiliation{Synergetic Innovation Center of Quantum Information and Quantum Physics,
University of Science and Technology of China, Hefei, 230026, China}
\date{\today}
\maketitle


\textbf{Probing collective spin dynamics is a current challenge in the field
of magnetic resonance spectroscopy and has important applications in
material analysis and quantum information protocols. Recently, the
rare-earth ion doped crystals are an attractive candidate for making
long-lived quantum memory. Further enhancement of its performance would
benefit from the direct knowledge on the dynamics of nuclear-spin bath in
the crystal. Here we detect the collective dynamics of nuclear-spin bath
located around the rare-earth ions in a crystal using dynamical decoupling
spectroscopy method. From the measured spectrum, we analyze the
configuration of the spin bath and characterize the flip-flop time between
two correlated nuclear spins in a long time scale ($\sim$ 1 s). Furthermore, we experimentally demonstrate that the rare-earth
ions can serve as a magnetic quantum sensor for external magnetic field.
These results suggest that the rare-earth ion is a useful probe for complex
spin dynamics in solids and enable quantum sensing in the low-frequency
regime, revealing promising possibilities for applications in diverse fields.%
}

\section*{Introduction}

Quantum memories are essential building blocks for a large-scale quantum
network \cite{SangouardRMP2011,NorthupNP2014} and can be constructed with
various physical systems \cite{SimonEPJD2010,HeshamiJMO2016}. There are
important evaluation criteria with relevance for practical applications, and
the key requirement hereby is the ability to store quantum states for
coherence times that are longer than the direct transmission time of the
channel \cite{SangouardRMP2011}. Strategies such as environmental and
materials engineering \cite{Sinclair2017,Balasubramanian2009}, quantum error
correction \cite{Chen2018} and optimal control pulses \cite{Genov2017} can
improve the coherence time, but, in general, the efficiency of these methods
is sensitive to the specific knowledge of the dynamics of environmental
fluctuations.

Rare-earth (RE) ion doped crystals have been recognized as a promising
candidate system for photonic quantum memory \cite%
{tittel,gisin,riedmatten,QD,faraon,ZhongN2015}. Storage of single photons
with extended lifetime \cite{gisin}, wide bandwidth \cite{tittel}, large
multimode capacity \cite{QD}, with telecom interface capability \cite{riedmatten} and in a nanoscale medium \cite{faraon} have
been demonstrated using this system. However, these remarkable results are
achieved using different samples and various experimental configurations.
Further combination of these techniques into a single system still remains a
significant challenge. In the RE doped crystal, the coherence of the RE ions
is affected by the magnetic fluctuations caused by surrounding nuclear
spins. Characterization of the environmental spectrum of the nuclear spin
bath will be particularly useful for identifying the optimal sample
configuration and design of novel memory protocols.

Dynamical decoupling (DD) spectroscopy has been employed as the useful tool
for detecting the spin dynamics in various systems \cite%
{BylanderNP2011,AlvarezPRL2011,YugePRL2011,BarGillNC2012,MalinowskiPRL2017}.
Pioneering work has focused on sensing individual nuclear spins and
nuclear-spin clusters in diamond \cite{KolkowitzPRL2012,ZhaoNN2012,ShiNP2014}%
, and resolving the bath of electronic spins on diamond surface \cite%
{RosskopfPRL2014,MyersPRL2014,RomachPRL2015}. Despite these remarkable
progress, probing the spin bath dynamics in photonic quantum memories,
especially in the low-frequency regime, is still unexplored. In this work,
we experimentally detect the collective dynamics of the Yttrium (Y) nuclear
spins in a Eu$^{3+}$ ions doped Y$_{2}$SiO$_{5}$ crystal using DD
spectroscopy method. To achieve the goal of understanding the bath dynamics,
we solve two important problems: characterization of the spectrum of bath
fluctuations (a double-Lorentzian spectrum), and determination of the
correlation time of spin bath (\textbf{$\sim $} $1$\textbf{\ }s). Using the
Eu$^{3+}$ ions as the probe, we further present a proof-of-principle
experiment on quantum sensing of an external magnetic field. These results
promise new applications in a wide range of fields from detection and
analysis of the complex many-body effects in solids to design of the
advanced quantum information processing protocols.

\begin{figure*}[tb]
\centering
\includegraphics[width=1.0\textwidth]{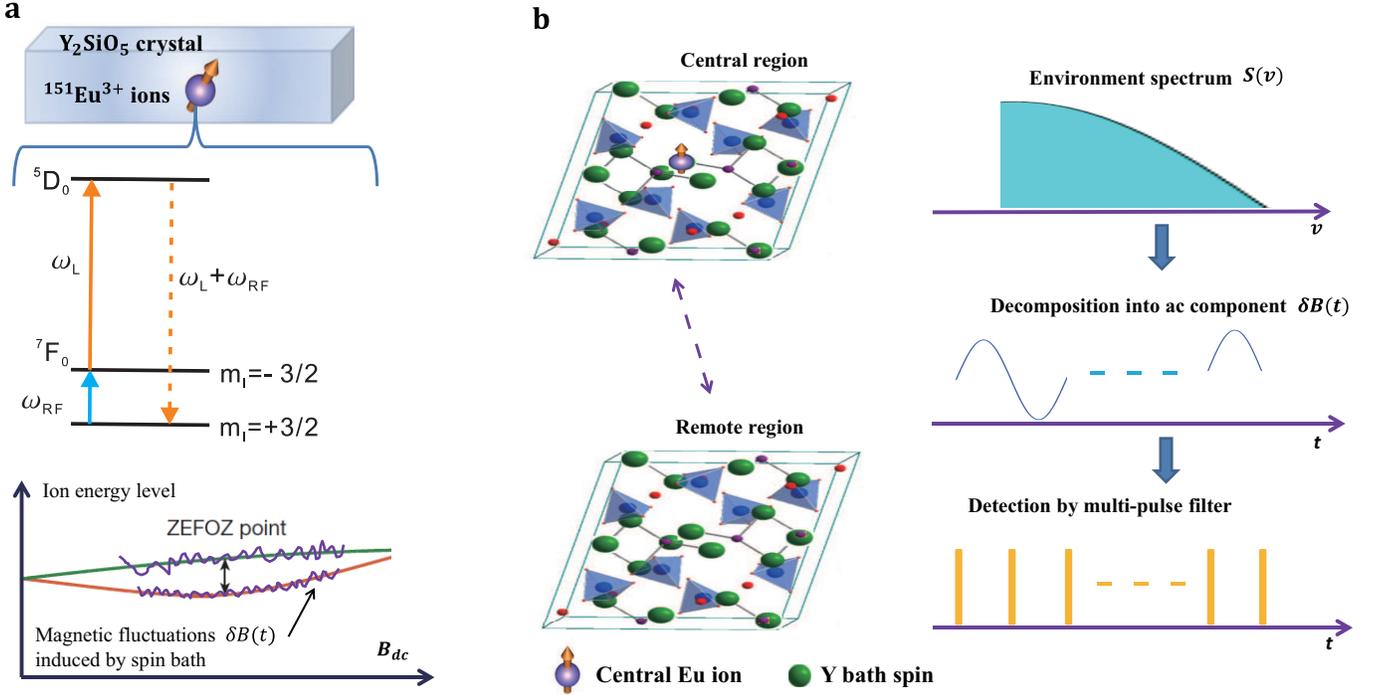}
\caption{\textbf{Schematic of the RE doped crystal and experimental scheme.}
\textbf{a}, Relevant level structure of the $^{151}$Eu$^{3+}$ ions in the Y$%
_{2}$SiO$_{5}$ crystal. We focus on the hyperfine transition between $%
|m_{I}=-3/2\rangle $ and $|m_{I}=+3/2\rangle $ in the $^{7}$F$_{0}$ electric
ground state of $^{151}$Eu$^{3+}$. The optical transition $^{7}$F$_{0}
\leftrightarrow ^{5}$D$_{0}$ is employed for the optical readout through
Raman-heterodyne detection \cite{ZhongN2015}. The dc magnetic field $B_{dc}$ is adjusted to
tune the energy level of the ion at different working points. The dynamics
of surrounding Y spin bath create magnetic fluctuation $\protect\delta B(t)$
around the RE ions. \textbf{b}, Illustration of the experimental method. We
measure the coherence of the $^{151}$Eu$^{3+}$ ions coupled to an
environment of Y nuclear spins. The whole system can be approximately
divided into two regions: the $^{151}$Eu$^{3+}$ ions with nearby Y bath
spins (central region) and the remaining Y bath spins (remote region). The
collective dynamics of spin bath induce a characterized environmental
spectrum, which can be decomposed into various frequency components (ac
magnetic field $\protect\delta B(t)$ with corresponding frequency $\protect%
\nu $). When a train of $\protect\pi $ pulses is applied synchronously with
this frequency, it acts as a filter to select this specific frequency
component (i.e., detect the ac magnetic field $\protect\delta B(t)$ with
frequency $\protect\nu $). Repeating the dynamical decoupling pulse sequence
for various pulse spacings $\protect\tau =1 /2\protect\nu $ yields the
environmental spectrum induced by the spin bath dynamics. }
\end{figure*}

\section*{Results}

\subsection*{Probing the spectrum of spin bath dynamics}

The physical platform is based on the hyperfine structure of the electronic
ground state ($^{7}$F$_{0}$) for $^{151}$Eu$^{3+}$ ions occupying site 1 Y
positions in the Y$_{2}$SiO$_{5}$ crystal. As illustrated in Fig. 1\textbf{a}%
, the doubly degenerate hyperfine state $\left\vert m_{I}=\pm \frac{3}{2}%
\right\rangle $ of $^{151}$Eu$^{3+}$ ion is split by the external magnetic
field field $B_{dc}$. Since the dynamics of nearby Y nuclear spins cause a
randomly fluctuating magnetic field $\delta B(t)$ at the $^{151}$Eu$^{3+}$
ion site, the perturbation contributing to detuning of the transition
frequency (i.e., decoherence between the energy levels) is \cite%
{hamiltonian,ZhongN2015}: $\omega _{RF}=\omega _{0}+S_{1}\delta
B(t)+S_{2}\delta B(t)^{2}(t)+...\mathrm{,}$ where $\omega_{0}$ is the
transition frequency at the bias field $B_{dc}$, $S_{1}$ ($S_{2}$) is the
first (second) order Zeeman coefficient (The complete characterization of
the spin Hamiltonian for $^{151}$Eu$^{3+}$ is given in Supplementary
Information). Since we are only interested in small perturbation field $%
\delta B(t)$ in this work, high orders Zeeman effect including $S_{2}$ can
be ignored. Thus the measured coherence time $T_{2}$ of the $^{151}$Eu$^{3+}$
ions can be used as an indicator of the Y spin bath dynamics $\frac{1}{T_{2}}%
\approx S_{1}\Delta B$, where $\Delta B$ is the variance of the magnetic
field fluctuations. When $B_{dc}$ approaches a critical magnetic field where
$S_{1}\simeq 0$, this is the so-called zero first order Zeeman (ZEFOZ) point
where $T_{2}$ of the considered spin transition will be greatly extended
\cite{ZhongN2015}.

We detect the spin bath dynamics by employing the spectroscopy method based
on DD \cite{BylanderNP2011,AlvarezPRL2011,YugePRL2011,BarGillNC2012}. The
experimental scheme are schematically shown in Fig. 1\textbf{b}. In the
intuitive picture, the collective dynamics of nuclear spin bath consists of
different kinds of components in the frequency domain. One component of the
spin bath dynamics induces an a.c. magnetic field fluctuation with a
characteristic frequency $v$. The application of multiple short $\pi $
pulses filters out the spin bath fluctuation $\delta B(t)$ at a narrow-band
range of frequency $v$ \cite%
{BylanderNP2011,AlvarezPRL2011,YugePRL2011,BarGillNC2012}:%
\begin{equation}
U(t)=\int_{0}^{\infty }F(vt)S(v)\text{d}v.
\end{equation}%
Here, $C=\exp (-U(t))$ corresponds to spin-echo signal amplitude of the $%
^{151}$Eu$^{3+}$ ions, $S(\nu )$ is the spectral density of the spin bath
fluctuation and $F(vt)$ is the filter function determined by the pulse
sequences (see Methods). This allows one to extract a wealth of information
about the environmental spectrum $S(v)$ from the measured spin-echo data and
hence probe the particular dynamics of spin bath.

\begin{figure}[]
\includegraphics[width=1.0\columnwidth]{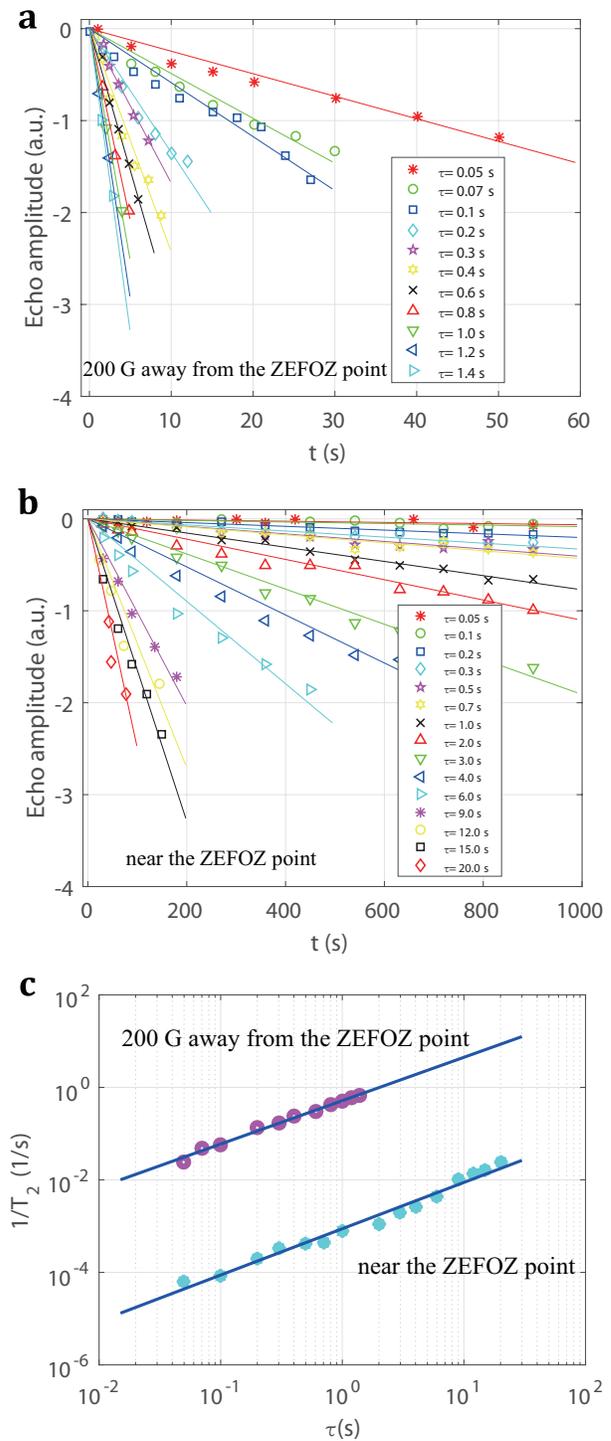}
\caption{\textbf{Coherence behaviors of the RE ions under dynamical decoupling control.} Adjusting the
alignment ${B}_{dc}$, the system can be placed in the case of (\textbf{a})
200 G away from ZEFOZ point and (\textbf{b}) near the ZEFOZ point with field
detuning $<$ 1G. As the CPMG pulse interval $\protect\tau $ increases, the
spin-echo signal (presented in log scale) decays with the total time $t$ and
decays increasingly quickly. The dots are measured data, and the curves are
fits to $\exp (-t/T_{2})$ to extract the coherence time $T_{2}$. \textbf{c},
The coherence time of the RE ions is proportional to pulse interval $%
\protect\tau $ with a scaling form $\frac{1}{T_{2}}\sim \protect\tau ^{%
\protect\beta }$ and $\protect\beta =1$.}
\end{figure}

Details about the experimental procedures are presented in Methods and
Supplementary Information. In short, optical pumping is firstly employed to
polarize the $^{151}$Eu$^{3+}$ ions into the $|m_{I}=+3/2\rangle $ spin
level. After radiofrequency (RF) multiple-pulse manipulation, the finally
emitted spin-echo signal is probed optically. The measured spin-echo
amplitude $C$ under the Carr-Purcell-Meiboom-Gill (CPMG) pulse sequence as a
function of pulse spacing $\tau $ is displayed in Fig. 2\textbf{a} and 2%
\textbf{b}. Here, we set the working condition as $B_{dc}$ being $200$ G
greater than the ZEFOZ field (Fig. 2\textbf{a}) and near the ZEFOZ point (Fig.
2\textbf{b}), respectively. The ZEFOZ field is chosen as approximately [$%
0.685$ T, $-0.812$ T, $0.714$ T] along the [$D1$, $D2$, $b$] crystal axis
\cite{ZhongN2015}. Then, we fit the experimental data with an exponential
form $\exp (-t/T_{2})$, from which we can extract the coherence time $T_{2}$
of the RE ion for different pulse intervals $\tau $, as shown in Fig. 2%
\textbf{c}. Diagram Fig. 2\textbf{c} shows several interesting features of
the coherence behaviors: First, the RE ion has much longer coherence times
with more applied pulses, which clearly illustrates the efficiency of the DD
pulse sequences for the present system. The coherence time of the RE ion is
measured to be $T_{2}=73.8$ s for pulse interval $\tau =12$ s, while it
extends significantly to $T_{2}=259 $ min ($4.3$ hours) for pulse interval $%
\tau =0.05$ s. Second, the decoherence rate can be described by a scaling
formula $\frac{1}{T_{2}}\sim \tau ^{\beta }$ with power $\beta =1$ (the
solid line in the Fig. 2\textbf{c}). A similar scaling has been observed in
the nitrogen-vacancy (NV) centers in diamond system \cite{deLangeS2010},
while in our case, the scaling spans over a several-hour-long time scale.
Moreover, the value of the power $\beta =1$, obviously deviates from the
number $\beta =2$ in the NV center system \cite{deLangeS2010}, which means
that there is a different mechanism at work. The robust scaling behaviors
provide evidence that the decoherence of central RE ions originates from the
intrinsic dynamics of the spin bath around the RE ions.

In Fig. 3\textbf{a}, we present the extracted spectrum of the spin bath
dynamics with $B_{dc}$ of 200G greater than ZEFOZ field. The typical
spectrum obtained is dominated by the contribution from the low-frequency
spin bath dynamics. We fit the detected result as a double-Lorentzian
spectrum $S(v)=\frac{1}{\pi }\frac{2b^{2}\tau _{c}^{s}}{v^{2}(\tau
_{c}^{s})^{2}+1}+\frac{1}{\pi }\frac{2b^{2}\tau _{c}^{f}}{v^{2}(\tau _{c}^{f})^{2}+1},$
where $b=2.46$ Hz is the average coupling strength of the bath to the
central RE ions, and $\tau _{c}$ is the correlation time of the bath
dynamics. In particular, the environmental spectrum is characterized by a
\textquotedblleft two-time-scale": $\tau _{c}^{s}=2.54$ s for the slow
dynamics component (see the green region in Fig. 3a), and $\tau
_{c}^{f}=0.25 $ s for the fast dynamics component (see the yellow region in
Fig. 3a).

\begin{figure}[tbh]
\includegraphics[width=1.0\columnwidth]{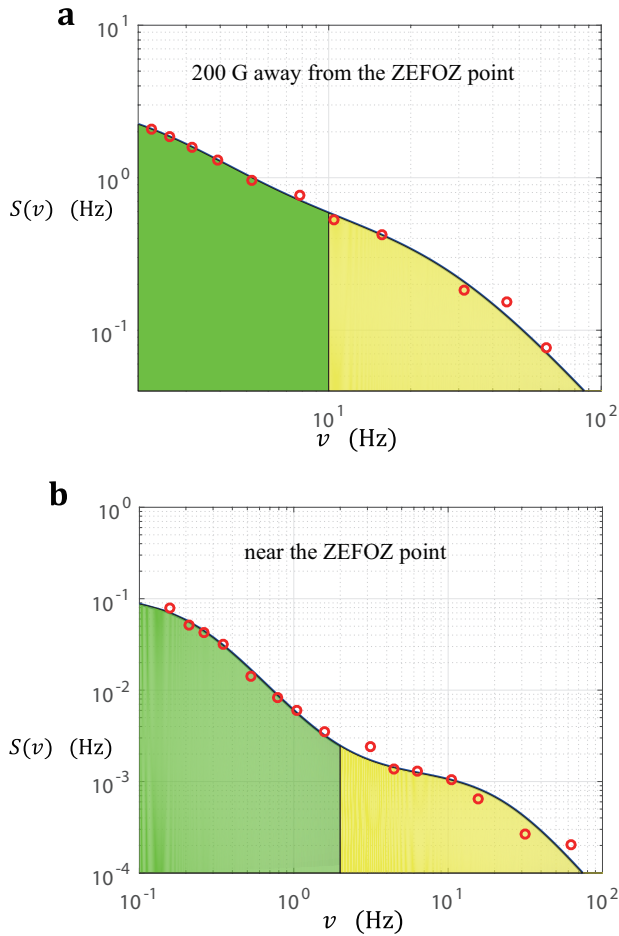}
\caption{\textbf{Extracted environmental spectrum through the dynamical
decoupling approach. }The detected spectrum $S(v)$ at 200 G away from the
ZEFOZ point and near the ZEFOZ point are shown in diagram (\textbf{a) }and (%
\textbf{b), }respectively. The measured spectrum data (red dots) are fitted
to a double-Lorentzian shape (blue line), where a strong contribution to the
spectral density comes from the slow dynamics component (green region) and
is accompanied by a weak contribution from the fast dynamics component
(yellow region).}
\end{figure}

In Fig. 3\textbf{b} we show the probed environmental spectrum with $B_{dc}$
near the ZEFOZ field. The measured spectrum are in overall similar to that
far away from the ZEFOZ case, confirming that the \textquotedblleft
two-time-scale" environmental dynamics are responsible and universal for the
present system. The characterized \textquotedblleft two-time-scale" values
are $\tau _{c}^{s}=28.6$ s for the slow dynamics component and $\tau
_{c}^{f}=0.25$ s for the fast dynamics component. In addition, the spectrum
have a much smaller amplitude of $b=0.12$ Hz, one order of magnitude smaller
than those far away from the ZEFOZ case. Comparing the data in Fig. 3\textbf{%
a}, we find that the dynamics strength of the bath spins is clearly
suppressed (i.e., slower $\tau _{c}$ and smaller $b$) near the ZEFOZ point.

\subsection*{Analysis of the spin bath dynamics}

Diagram Fig. 3 shows several remarkable features of the collective spin bath
dynamics: First, we find that the typical environmental spectra in solids
such as power law spectra $\frac{1}{v^{\alpha }}$ in a superconducting flux
qubit \cite{BylanderNP2011,QuintanaPRL2017} and simple Lorentzian spectra $%
\frac{1}{v^{2}\tau _{c}^{2}+1}$ in bulk diamond systems with various NV
center densities and impurity concentrations \cite{BarGillNC2012}, cannot
describe our measured data (the fitting is rather poor, please see\textbf{\ }%
Fig. S2 in the Supplementary Information). This suggests different
environmental dynamics in the present system compared to that in other
solids. Second, the environment spectrum at different working points have a
similar fast correlation time $\tau _{c}^{f}=0.25$ s, while they have
different slow correlation time $\tau _{c}^{s}$. Under the same
multiple-pulse control, the RE ions can achieve different coherence time $%
T_{2}$ for different slow correlation time $\tau _{c}^{s}$: for comparison,
the maximum $T_{2}=40$ s for $\tau _{c}^{s}=2.54$ s (200G away from the
ZEFOZ point), the maximum $T_{2}=$ $259$ min for $\tau _{c}^{s}=28.6$ s
(near the ZEFOZ point). We also note that the observed characteristic time $%
\tau _{c}$ is orders of magnitude longer than that of other candidate system
for quantum memories \cite{deLangeS2010}, which suggests that the probed
spin bath dynamics are in the low-frequency regime.

We initially explain this double-Lorentzian spectra is attributed to the
special dynamics of the spin bath. The dynamics of Y spin bath include
single nuclear-spin precession and flip-flop processes of nuclear-spin
pairs. Our experiments are performed in strong magnetic fields, where the
single nuclear-spin precession is strongly suppressed. Thus, the dynamics of
the spin bath are dominated by the interacting nuclear-spin pairs. When one
bath spin \thinspace $i\,$flips, a corresponding bath spin $j$ flops in the
opposite direction. A large number of flip-flop events show that the
configuration of spin bath changes over time, which can be considered as the
collective dynamics of spin bath with different kinds of flip-flop rates $%
R_{ij}$.

Moreover, the strong coupling between the central $^{151}$Eu$^{3+}$ ions and
neighboring Y bath spins induces that Larmor frequencies of these bath spins
are detuned from others and cannot readily exchange. As shown in Fig. 1b,
this create a so-called ``frozen core" region \cite{ZhongN2015}, where the
flip-flop rate $R_{ij}$ of these bath spins is significantly suppressed.
Therefore, the dynamics of the bath spins can be characterized by two
distinct time scales: slow dynamics $\tau _{c}^{s}\sim $ $1/R^{s}$ (small
flip-flop rate $R^{s}$) in the ``frozen core" region and fast dynamics $\tau
_{f}^{s}\sim $ $1/R^{f}$ (large rate $R^{f}$) outside the ``frozen core"
region, respectively.

The flip-flop events show that the configuration of the spin bath changes in
time. As a result, the bath dynamics create a fluctuating magnetic field $%
\delta B(t)\sim \sum_{i}n_{i}(t)\,$\ as a sum of a large number of
independent configurations $\{...n_{i}(t),...n_{j}(t),...\}$, where $%
n_{i}(t)=1$ if $i$-th bath spin at time $t$ is $\left\vert
n_{i}\right\rangle =\left\vert \uparrow \right\rangle $, and $n_{i}(t)=-1$
for $\left\vert n_{i}\right\rangle =\left\vert \downarrow \right\rangle $.
Thus the environmental dynamics can be modeled as a random process with
Gaussian distribution according to the central limit theorem \cite%
{deLangeS2010}. Due to lack of coherence between different configurations of
spin bath, this dynamics also does not depend on history, i.e. it is
Markovian. Summarizing all of the above, we can expect the environment
dynamics in each region can be represented as a correlation function $%
\left\langle \delta B(t)\delta B(0)\right\rangle \sim \exp (-t/\tau _{c})$
\cite{DobrovitskiPRL2009,deLangeS2010}. Here $\tau _{c}=1/R$ is the
correlation time of this process. Since there are two regions with different
time scales or flip-flop rates $\tau _{c}^{s}=1/R^{s}$, $\tau
_{c}^{f}=1/R^{f}$, we conclude that there is a double-Lorentzian spectrum
consisting of two components $S(v)\sim \frac{1}{v^{2}(\tau _{c}^{s})^{2}+1}+%
\frac{1}{v^{2}(\tau _{c}^{f})^{2}+1}$ (More quantitative analysis are
provided in the Supplementary Information).

\subsection*{Detection of an external magnetic field}

The above results demonstrate that the RE ions can be used for detecting the
spin-bath spectrum inside the crystal. Inspired by this experimental
precursor, next we report that the RE ions can also be employed for
detection of an external field. An example measurement for an external ac
magnetic field $B_{ac}(t)$ is presented in Fig. 4. Details about the experimental procedures are presented in the Supplementary Information.
Here, we set the working
condition as $B_{dc}$ being $200$ G greater than the ZEFOZ field. Simple
two-pulse spin-echo sequence is employed for sensing an synchronized
external field \cite{MazeN2008}. The accumulated phase during the two-pulse
sequence can be written as $\phi =\int_{0}^{\tau
}S_{1}B_{ac}(t)dt-\int_{\tau }^{2\tau }S_{1}B_{ac}(t)dt=\pi \tau S_{1}B_{ac}$%
, where $2\tau $ is the total evolution time and $B_{ac}(t)=B_{ac}\sin (2\pi
vt)$, with frequency $\nu =1/(2\tau )$, is the external field to be
measured. Thus, detecting the relative phase shift $\phi $ induced by the
magnetic field enables precise determination of the applied ac
magnetic field $B_{ac}(t)$. The real part ($X$) and the imaginary part ($Y$) of the
spin-echo signal are measured simultaneously to minimize the affects from
the fluctuations of the laser power and RF power (Detailed discussions are given
in the Supplementary Information). As shown in Fig. 4a, when the amplitude
of the ac magnetic field increases, due to the accumulation of the phase,
the spin-echo signal varies periodically. A sensitivity of $10.8$ nT/$\sqrt{%
\mathrm{Hz}}$ is achieved at a detecting frequency of $0.75$ Hz. Another
example measurement performed with a working condition of $B_{dc}$ being 6 G
greater than the ZEFOZ field is presented in Fig. S4 in the Supplementary
Information. We achieved a sensitivity of $118$ nT/$\sqrt{\mathrm{Hz}}$ at a
detecting frequency of $66$ mHz. Therefore, the present RE ion magnetometer
enables sensitive detection of external fields in the low-frequency regime near $0.1$ Hz, which
is not accessible by other quantum sensors with much shorter coherence times
\cite{DegenRMP2017}.

\begin{figure}[tbh]
\includegraphics[width=1.0\columnwidth]{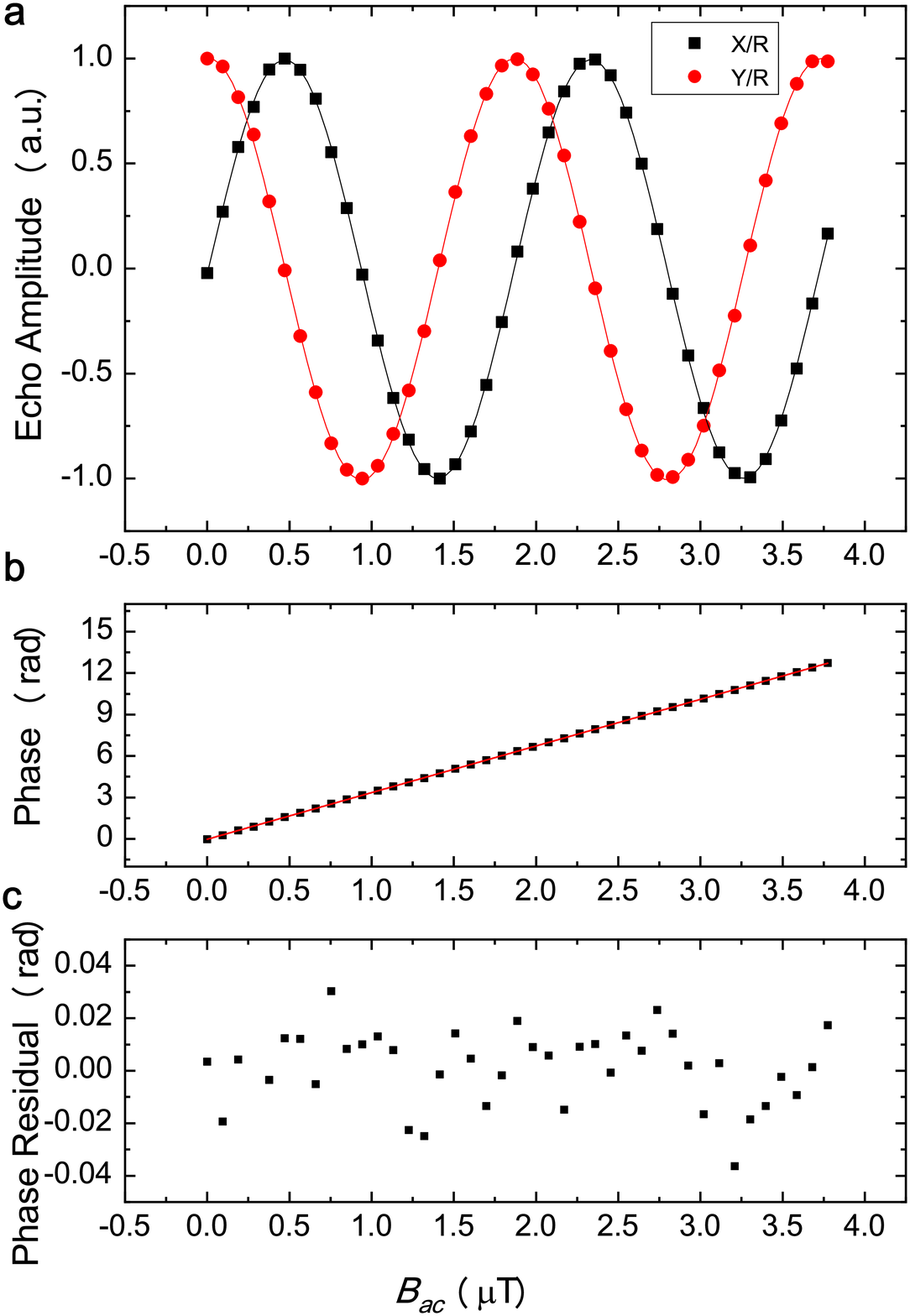}
\caption{\textbf{Demonstration of quantum sensing of an external ac magnetic
field.} \textbf{a}, Examples of measured spin-echo signal as a function of $%
B_{ac}$. The working condition is set as $B_{dc}$ being 200 G greater than
the ZEFOZ field where the $T_2$ for the sensor ions is approximately 1.44 s.
The evolution time for the spin-echo is $2\protect\tau =1.332s$, which
corresponds to an operating frequency of $0.75$ Hz. Each point is the
average of four measurements with total measurement time of 5.328s. The
black squares and red dots correspond to $X/R$ and $Y/R$, where $X$, $Y$, $R$
are the real part, imaginary part and amplitude of the spin-echo signal,
respectively. The black line and red line are the sinusoidal fit for the experimental
data. \textbf{b}, The phase accumulation for the spin echo. $\protect\phi %
=\arctan ({X/Y)}$. A linear fit (red line) for the data produces a phase response of $%
3.376\pm0.003$ $rad$/$\protect\mu$T. \textbf{(c)} The fit residuals for each
measurement. The standard error for the phase measurement $\protect\delta
\protect\phi $ is $0.016$ $rad$, which translates into a minimum detectable
magnetic field $\protect\delta B_{min}$ of $4.7$ nT.}
\end{figure}

\section*{Discussion}

In conclusion, we have measured and characterized the collective dynamics of
nuclear-spin bath in a RE ion doped crystal. From the coherence measurements
of the central  $^{151}$Eu$^{3+}$ ions under multiple-pulse control and various
magnetic fields, we determine the spectrum of the Y bath dynamics and its
correlation times. The RE ions doped in solids can also serve as a quantum
sensor for external magnetic fields. With this observation, this
proof-of-principle work could be the first step towards the detection of
various environmental processes in other RE doped solids, especially in the
Er$^{3+}$: Y$_{2}$SiO$_{5}$ crystal, a candidate for a quantum memory
working in the $1550$\thinspace nm communication band \cite{Er}. Recent
investigations have also facilitate the detection and manipulation of single
RE ion in a nanocrystal with long coherence time \cite%
{UtikalNC2014,SiyushevNC2014}. Hence, this work also opens up wide
perspective for applications in the field of detecting the interesting
dynamics of complex materials with high-sensitivity, in the low-frequency
regime and with nano-scale resolution.

\section*{Methods}

\subsection*{Experimental setup}

The sample is a $0.01\%$ doped Eu$^{3+}$:Y$_{2}$SO$_{5}$ crystal with a
thickness of $1$ mm along the crystal's $b$-axis. The sample is oriented
within a superconducting magnet contained in a cryogen-free cryostat and its
temperature is maintained at $1.80\pm0.01$ K. The $580$-nm laser beam from a
frequency-doubled semiconductor laser is further modulated with an
acoustic-optic modulator to generate desired preparation and probe light.
The RF field, which is controlled by an arbitrary waveform generator and
amplified by a $300$-W amplifier, is employed to stimulate different RF
pulse schemes.

The RF spin-echo signal is readout by optical Raman heterodyne
detection \cite{ZhongN2015}. Before the RF manipulation sequence, the pump
laser prepares the large population difference in the considered spin
transition to obtain a large signal. After the RF manipulation, a
synchronized laser pulse at $\omega _{L}$ ($\sim 517.471$ THz) converts the
RF spin-echo into an optical pulse at $\omega _{L}+\omega _{RF}$ through a
two-photon process and Raman scattering. The spin echo is observed by beating the generated light at $\omega _{L}+\omega _{RF}$ and the probe light at $\omega _{L}$.
The phase accumulation of the spin-echo is observed by mixing the beat signal of the spin-echo with two
separate RF local oscillator channels separated in phase by $90^{\circ }$.
More details about the experimental setup can be found in the Supplementary
Information.

\subsection*{Spectroscopy method based on dynamical decoupling control}

We quantify the coherence of central RE ions by using the function $C=\exp
[-U(t)]$, which is defined as the off-diagonal elements of the density
matrix of the spin state of RE ions. Formally, the dynamics of the RE spin
are analyzed using the evolution operator. On the one hand, the decoherence
process from the environmental fluctuation $\delta B(t)$ is introduced by $%
\exp (-\frac{i\sigma _{z}\delta B(t)}{2}t)$. On the other hand, DD sequences can be considered as applying a certain number of $\exp
(-\frac{i\sigma _{x}\pi }{2})=-i\sigma _{x}$ at time $t_{i}$. Therefore, we
can write the evolution operator by slicing the time interval into small
pieces:
\begin{equation*}
C(t)=\left\langle \exp (-i\int_{0}^{t}F(t^{^{/}})\delta
B(t^{^{/}})dt^{^{/}})\right\rangle ,
\end{equation*}%
where $F(t)$ is the temporal filter function, which takes value $-1$ and $+1$
and changes the sign at each pulse time $t_{i}$. Then, using the Fourier
transformation of the above equation, $U(t)$ can be calculated through the
spectral density $S(v)$ of the environmental fluctuation as Eq. (1).

The filter function for the $n$-pulse CPMG control is calculated as $%
F(vt)=8\sin ^{2}(vt)\sin ^{4}(vt/4n)/\cos ^{2}(vt/2n)$ \cite{CywinskiPRB2008}%
. Since we employ the pulse sequences with a large number $n$, the filter
function $F(vt)$ can be effectively approximated as a $\delta $-function in
the frequency domain. Flipping the central RE ions by a sequence of $\pi $
pulses will suppress the effect of the fluctuating magnetic field components
at every other frequency while filtering the fluctuating magnetic filed
component oscillating synchronously with the pulse interval $\tau =1/2v$.
Thus, repeating the experiment for different pulse spacings $\tau $, we can
scan the entire spectrum $S(v)$ of the environment.



\acknowledgments\textbf{Acknowledgments} This work was supported by the
National Key R\&D Program of China (No. 2017YFA0304100), the National
Natural Science Foundation of China (Nos. 61327901, 11774331, 11774335,
11474270, 11504362, 11654002), Anhui Initiative in Quantum Information Technologies (No. AHY020100). Key Research Program of Frontier Sciences,
CAS (No. QYZDY-SSW-SLH003), the Fundamental Research Funds for the Central
Universities (Nos. WK2470000023, WK2470000026).

\textbf{Author Contributions} Z.-Q. Z., T. T. and C.-F. L. designed
experiment. Z.-Q. Z. carried out the experiment assisted by Y. M., P.-Y. L,
Z.-F. L., C. L., P.-J. L., X. L., Y.-L. H., T.-S. Y., J. H., X. L., and
Y.-X. X. Z.-Q. Z., T. T. and Y. M. analyzed the experimental results. T. T.
analyzed the dynamical decoupling spectroscopy. Y. M. and Y.-J. H. did the
calculation for the spin Hamiltonian. C.-F. L. and G.-C. G. advised on the
project. Z.-Q. Z. and T. T. wrote the paper with input from other authors.
All authors discussed the experimental procedures and results.

\textbf{Competing financial interests} The authors declare no competing
financial interests.

\end{document}